\documentclass[11pt]{article}
\usepackage{moriond,epsfig}
\def\be{\begin{equation}}
\def\ee{\end{equation}}
\def\bea{\begin{eqnarray}}
\def\eea{\end{eqnarray}}

\begin{document}
\vspace*{4cm}
\title{NEW CONSTRAINTS ON VARYING ALPHA}

\author{ C.J.A.P. MARTINS }

\address{CFP, R. do Campo Alegre 687, 4169-007 Porto, Portugal,\\
and Department of Applied Mathematics and Theoretical Physics, CMS,\\
University of Cambridge, Wilberforce Road, Cambridge CB3 0WA, UK}

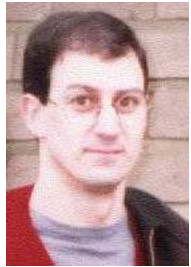
\maketitle\abstracts{
I briefly present some current theoretical motivations for
time or space variations of the fundamental constants of nature and review
current key observational results. I focus on the fine-structure constant,
and particularly on measurements using quasars and the cosmic microwave background. I also compare various observational results to the simplest
model-building expectations.}

\section{Introduction}

One of the most valued guiding principles (or should one say beliefs?)
in science is that there ought to be a single, immutable set of laws
governing the universe, and that these laws should remain the same
everywhere and at all times. In fact, this is often generalised into
a belief of immutability of the universe itself---a much stronger
statement which doesn't follow from the former.
It was Einstein (who originally introduced the cosmological constant as a
`quick-fix' to preserve a static universe) who
taught us that space and time are not an immutable arena
in which the cosmic drama is acted out, but are in fact part of the
cast---part of the physical universe. As physical entities, the
properties of time and
space can change as a result of gravitational processes.
Interestingly enough, it was soon after the appearance of General Relativity,
the Friedman models, and Hubble's discovery of the expansion of the
universe---which shattered the notion of immutability of the
universe---that time-varying fundamental constants
first appeared in the context of a complete cosmological model\cite{1}.

Despite the best efforts of a few outstanding theorists, it took as usual
some observational claims of varying fundamental constants\cite{2}
to make the alarm bells sound in the community as a whole,
and start convincing previously sworn skeptics. In the past
two years there has been an unprecedented explosion of interest in this area,
perhaps even larger than the one caused a few years ago by the evidence
for an accelerating universe provided by Type Ia supernova data.
Observers and experimentalists have tried to reproduce these results
and update and improve existing constraints, while a swarm
of theorists has flooded scientific journals with a range
of possible explanations---for reviews see\cite{roy,book,uzan}.

The so-called fundamental constants of nature are widely regarded as
some kind of distillation of physics.
Their units are intimately related to the form and structure of
physical laws. Despite their perceived fundamental nature, there is
no theory of constants as such.
One common view of constants is as asymptotic states. For example, the
speed of light $c$ is (in special relativity) the maximum velocity
of a massive particle moving in flat spacetime. The gravitational
constant $G$ defines the limiting potential for a mass that doesn't form
a black hole in curved spacetime. The reduced Planck constant
$\hbar\equiv h/2\pi$ is the universal quantum of action and hence defines a
minimum uncertainty. Similarly in string theory there
is a fundamental unit of length, the characteristic size of the strings.
So for any physical theory we know of, there should be one such constant.
This view is acceptable in practice, but unsatisfactory in principle,
because it doesn't address the question of the constants' origin.

Another view is that they are simply necessary (or should one say
convenient?) inventions: they are not really fundamental
but simply ways of relating quantities
of different dimensional types. In other words, they are simply conversion
constants .This view, first clearly formulated
by Eddington\cite{4} is at the origin
of the tradition of absorbing constants in the equations of physics.
However, it should be remembered that this procedure
can not be carried arbitrarily far.
For example, we can consistently set $G=h=c=1$, but we cannot set
$e=\hbar=c=1$ ($e$ being the electron charge)
since then the fine-structure constant would have the value
$\alpha\equiv e^2/(\hbar c)=1$ whereas in the real world $\alpha\sim1/137$.

Perhaps the key point is the one made in\cite{5}:
for example, if there was no fundamental length, the properties of physical
systems would be invariant under an overall rescaling of their size,
so atoms would not have a characteristic size, and we wouldn't even be able
to agree on which unit to use as a `metre'. With a fundamental quantum unit of
length, we can meaningfully talk about short or large distances.
In other words, `fundamental' constants are fundamental only
to the extent that they provide us with a way of transforming any quantity
(in whatever units we have chosen to measure it) into a pure number whose
physical meaning is immediately clear and unambiguous.

It is believed that the unification of the known fundamental interactions of
nature requires theories with additional spacetime dimensions. Even though
there are at present no robust ideas about how one can go from these theories
to our familiar low-energy 4D spacetime,
it is clear that such a process will necessarily involve procedures known
as dimensional reduction and compactification, with the consequence
that the ordinary 4D constants become `effective' quantities,
typically being related to the true higher-dimensional fundamental
constants through the characteristic length scales of the
extra dimensions. It also happens that these length scales typically have a
non-trivial evolution. 
In these circumstances, one is naturally led to the expectation of
time and space variations of the 4D constants we can measure.
In what follows we will focus on the fine-structure constant
($\alpha\equiv e^2/\hbar c$, a measure of the strength of electromagnetic
interactions). Although it is a path that will not be pursued in detail here,
it is important to
keep in mind that the search for varying constants has crucial relevance
in the context of tests of the behaviour of gravity\cite{7}.

\section{Local Experiments}

Laboratory measurements of the value of the fine-structure constant,
and hence limits on its variation, have been carried out for a
number of years.
The best currently available limit is by the Paris (BNM-SYRTE) group\cite{9}
\begin{equation}
\frac{d}{dt}{\ln{\alpha}}<1.2\times 10^{-15}\,{\rm yr}^{-1}\,,
\label{labbound}
\end{equation}
on a timescale of 57 months.Note that this bound is local (it applies to the present day only).
This is obtained by comparing rates between atomic clocks
(based on ground state hyperfine transitions) in alkali atoms with different
atomic number $Z$.
The current best method uses ${}^{87}Rb$ vs. ${}^{133}Cs$ clocks,
the effect being a relativistic correction of order $(\alpha Z)^2$.
Outstanding progress is being made in this area:
laser-cooled, single-atom optical
clocks and (more importantly) performing such experiments in space, is
expected to improve these bounds (as well as those on Equivalence Principle tests) by several orders of magnitude.
Roughly speaking, ACES (a French experiment to be carried out at the ISS)
should provide a one order of magnitude improvement, $\mu$SCOPE (a CNES satellite, scheduled for launch in 2007) can provide two, GG (an Italian mission) three, and STEP (a joint ESA-NASA mission) up to five orders of magnitude improvement on current bounds.

The best geophysical constraint comes
from the analysis of $Sm$ isotope
ratios from the natural nuclear reactor at the Oklo (Gabon) uranium mine,
on a timescale of $1.8\times 10^9$ years, corresponding to a cosmological
redshift of $z\sim0.14$. A recent analysis\cite{10} finds two possible ranges
of resonance energy shifts, corresponding to the following rates
\begin{equation}
\frac{\dot \alpha}{\alpha}=(0.4\pm0.5)\times 10^{-17}\,{\rm yr}^{-1},\quad
\frac{\Delta \alpha}{\alpha}=(-0.8\pm1.0)\times 10^{-8}
\label{oklobound1}
\end{equation}
\begin{equation}
\frac{\dot \alpha}{\alpha}=-(4.4\pm0.4)\times 10^{-17}\,{\rm yr}^{-1},\quad
\frac{\Delta \alpha}{\alpha}=(8.8\pm0.7)\times 10^{-8}\,.
\label{oklobound2}
\end{equation}
The authors also point out that there is tentative evidence that
the second result can be excluded by further $Gd$ and $Cd$ sample.
However, that analysis procedure is subject to more uncertainties (such as
sample contamination) than the one for $Sm$, so a more detailed analysis is
required before definite conclusions can be drawn.
One of the criticisms of the above work has to do with the fact that it assumes
a Maxwell-Boltzmann low energy neutron spectrum. A recent\cite{Lam} and
arguably more realistic treatment finds
\begin{equation}
\frac{\Delta \alpha}{\alpha}=(4.5\pm1.1)\times 10^{-8}\,, z\sim0.14\,.
\label{oklobound3}
\end{equation}
It should also be emphasised that these measurements are not 'clean', in
the sense that assumptions on the behaviour of other couplings must be made in
order to extract a value for $\alpha$. Finally, constraints can also be
obtained from Rhenium abundances in meteorites. However, these are thought to
be less reliable, so we will not discuss them here.

\section{The Recent Universe}

The standard technique for this
type of measurements, which have been attempted
since the late 1950's, consists of observing the fine splitting of
alkali doublet absorption lines in quasar spectra, and comparing these with
standard laboratory spectra. A different value of $\alpha$ at early times would
mean that electrons would be more loosely (or tightly, depending on the
sign of the variation) bound to the nuclei
compared to the present day, thus changing the
characteristic wavelength of light emitted and absorbed by atoms.
The current best result is\cite{11}
\begin{equation}
\frac{\Delta\alpha}{\alpha}=(-0.5\pm1.3)\times 10^{-5}\,.
\qquad z\sim2-3\,;
\label{varshbound1}
\end{equation}
Note that in comparing a rate of change at a certain epoch
(${\dot\alpha}/\alpha$) with a relative change over a certain range
(${\Delta\alpha}/\alpha$) one must choose not only a timescale (in order
to fix a Hubble time) but also a full cosmological model, in particular
specifying how $\alpha$ varies with cosmological time (or redshift). Hence
any such comparison will necessarily be model-dependent.

Recent progress has focused on a new technique, commonly called the {\it Many
Multiplet} method, which uses various multiplets from many
chemical elements to improve the accuracy by about an order of
magnitude. The current best result is\cite{12} 
\begin{equation}
\frac{\Delta\alpha}{\alpha}=(-0.54\pm0.12)\times 10^{-5}\,, \qquad z\sim0.2-3.7\,,
\label{varshbound6}
\end{equation}
corresponding to a 4.7-sigma detection of a {\it smaller} $\alpha$
in the past. This comes from data of 128 quasar absorption
sources from the Keck/HIRES telescope, and despite extensive testing
no systematic error has been found that could explain the result. On the
other hand, a more recent analysis\cite{13} using only 23 sources of VLT/UVES
data and a less thorough analysis finds either
\begin{equation}
\frac{\Delta\alpha}{\alpha}=(-0.06\pm0.06)\times 10^{-5}\,, \qquad z\sim0.4-2.3\,
\label{varspetit}
\end{equation}
assuming today's isotopic abundances or, if one instead assumes
isotopic abundances typical of low metalicity, a detection,
\begin{equation}
\frac{\Delta\alpha}{\alpha}=(-0.36\pm0.06)\times 10^{-5}\,, \qquad z\sim0.4-2.3\,;
\label{varspetit2}
\end{equation}
while the true value should be somewhere between the two, these results
also highlight the need for independent checks.

A different approach is using radio and millimetre spectra of
quasar absorption lines. Unfortunately at the moment this can only be
used at lower redshifts, yielding the upper limit\cite{14}
\begin{equation}
\left|\frac{\Delta\alpha}{\alpha}\right|<0.85\times 10^{-5}\,, \qquad z\sim0.25-0.68\,.
\label{varshbound6a}
\end{equation}
Absorption lines can also be used to search for variations of
other dimensionless constants, such as
the proton-to-electron mass ratio ($\mu=m_p/m_e$), which
can be measured via the wavelengths of $H_2$ transitions in damped
Lyman-$\alpha$ systems (using the fact that electron
vibro-rotational lines depend on the reduced mass of the molecule, the
dependence being different for different transitions). This method has
produced another claimed\cite{15} 3-sigma detection\footnote{This number
has been recently revised\cite{newmu} to $\frac{\Delta\mu}{\mu}=(2.97\pm0.74)\times 10^{-5}$.}
\begin{equation}
\frac{\Delta\mu}{\mu}=(5.02\pm1.82)\times 10^{-5}\,,
\label{varshbound7}
\end{equation}
which is based on VLT/UVES data from a single quasar at $z\sim3$.
Note that if $\alpha$ varies so should $\mu$. The exact
relation between the two is unknown, but can be predicted for particular models.
When better quality data becomes available over a range of
redshifts these consistency relations will provide extremely constraining
tests.

A further alternative is to use {\it emission} rather than absorption lines.
Indeed, the first astrophysical measurements of $\alpha$ used emission lines,
though soon absorption became preferred. It is interesting that there have
been almost no high-redshift measurements of $\alpha$ using emission lines for
over three decades. Recent\cite{Bahcall} measurements have been obtained
from strong OIII emission lines from quasars in the SDSS. A dataset of 165
spectra at $0.16<z<0.80$ yielded
\begin{equation}
\frac{\Delta\alpha}{\alpha}=(1.2\pm0.7)\times 10^{-4}\,.
\label{bahc}
\end{equation}
Compared to absorption, the emission method is quite simple and
straightforward in principle, though possibly less sensitive and harder to
apply to higher redshifts. 

We\cite{jarle} have obtained
data in service mode at the VLT/ISAAC during period 72, in order to carry
out a pilot study for an improved emission-based method. Our sources are
in the range $2<z<3$, and a preliminary analysis seems to indicate
values of $\alpha$ that are larger in the past. While various issues in
our pipeline require further analysis (notably wavelength calibration and noise), it is interesting to speculate on the observation that absorption
methods seem to prefer smaller values of $\alpha$ in the past, while emission
methods seem to prefer larger values. Of course the simplest explanation is
perhaps systematic errors---though a very interesting question is whether
there is some {\it relative} systematics, meaning some reason why emission
and absorption methods would give different answers {\it even if there is
no $\alpha$ variation at all}. On the other hand, the optimist's view would
be that we might be starting to probe spatial variations of $\alpha$.
Indeed, emission and absorption methods measure $\alpha$ in very different
physical environments. If one is willing to buy the idea that $\alpha$ is
not a constant in the first place, then there is no real reason to expect that
it should have the same value in both environments. Of course, if this is
the case it will be quite difficult to distinguish there variations from
other effects

\section{Interlude: Does This Make Sense?}

Even though comparing constraints at different
epochs is subject to a number of {\it caveats}, it is clear that not all the
above data is consistent, so it is important to reflect on the current
situation and its implications for model building. One is reminded of the
comment that {\it any model that fits all the data at a given time is
necessarily wrong, because at any given time not all the data are correct.}
This means that at the moment it is absolutely pointless to introduce new toy
models with five additional free parameters which if tuned to three decimal
places can adequately explain all observations. Chances are that if/when they
appear in print such models will already be ruled out.

It is obvious that a measurement of say $\alpha(z)$ will be of
fundamental importance, but also clear that the current data
is far from adequate for this purpose. In these circumstances, we take the view
that the purpose of models is not to fit the data but to sharpen the questions.
We\cite{mods} have therefore studied the simplest class of varying
$\alpha$ models (where the variation is due to a coupling of a scalar field
to the electromagnetic field tensor) and considered the particular case where
the model's free functions (potential and gauge kinetic function) are
Taylor-expanded up to linear order. Note that any realistic model of this
type reduces to such a model for some time interval around to day. it is also
assumed that the scalar field providing a varying $\alpha$ also provides the
dark energy.

The distinguishing characteristic of the linearised models is that the
evolution of $\alpha$ is quite significant at very low redshifts. Consequently,
it is not possible to reconcile the Oklo and Webb results in the context of
these models. If we take the Oklo data seriously, the maximum $\alpha$
variation allowed at $z\sim2$ is only about $10^{-7}$. Model parameters can
of course be chosen to be approximately consistent with all quasar data,
but they will the be inconsistent with Oklo. Of course, one can also entertain
the possibility that all observations are at least roughly right---in that
case the consequence would be that our linearity assumption must break
down on a timescale significantly smaller than a Hubble time (which may be
a symptom of fine-tuning). Given that a scalar field that produces a varying
$\alpha$ can also make a significant contribution towards the dark energy
of the universe, it is interesting to speculate on the possible relation between
the results of this analysis and hints for a time-varying equation of state
of dark energy.

\section{The Early Universe: BBN and CMB}

If variations exist in recent times, one is naturally led to the
question of what was happening at earlier times---presumably the
variations relative to the present day values would have been stronger.
These measurements\cite{me1,me2,me3,me4,me5,me6} are therefore crucial as an
independent check.
Two of the pillars of standard cosmology are noteworthy here.
Firstly, nucleosynthesis has the obvious advantage of probing
the highest redshifts ($z\sim10^{10}$), but
it has a strong drawback in that one is always forced to make
an assumption on how the neutron to proton mass difference depends on $\alpha$.
No known particle physics model provides such a relation, so one usually
has to resort to a phenomenological expression, which
is needed to estimate the effect of a varying $\alpha$ on the ${}^4He$
abundance. The abundances of the other light elements depend much less
strongly on this assumption, but on the other hand these abundances are
much less well known observationally.

The cosmic microwave background probes intermediate redshifts ($z\sim10^3$),
but allows model-independent measurements and has the significant advantage
that one has highly accurate data.
A varying $\alpha$ changes the ionisation history of the
universe: it changes the Thomson scattering cross
section for all interacting species, and also the recombination
history of Hydrogen (by far the dominant contribution) and other
species through changes in the energy levels and binding energies. This
has important effects on the CMB angular power spectrum. Suppose
that $\alpha$ was larger at the epoch of recombination. Then the position of
the first Doppler peak would move smaller angular scales, its amplitude would
increase due to a larger early Integrated
Sachs-Wolfe (ISW) effect, and the damping at small
angular scales would decrease.

We have recently carried out detailed analyses of the effects of a
varying $\alpha$ on BBN and the CMB, and compared
the results with the latest available observational results in each case. 
We find that although the current data has a very slight preference
for a smaller value of $\alpha$ in the past, it is consistent with
no variation (at the 2-sigma level) and, furthermore,
restricts any such variation, from the
epoch of recombination to the present day, to be less than a few percent.
Specifically, for BBN we find\cite{me3}
\begin{equation}
\frac{\Delta\alpha}{\alpha}=(-7\pm9)\times 10^{-3}\,, \qquad z\sim10^{10}\,,
\label{varsbbn}
\end{equation}
while for the CMB we have\cite{me5}
\begin{equation}
0.95<\frac{\alpha_{cmb}}{\alpha_{now}}<1.02\,, \qquad z\sim10^3\,.
\label{varscmb}
\end{equation}
The latter uses the WMAP first year data.\footnote{We eagerly await the release
of the WMAP second-year data, which will allow us to carry out a number of further tests that are currently not possible.}
In both cases the likelihood distribution function is skewed towards smaller
values of $\alpha$ in the past.
One of the difficulties with these measurements is that one needs to find
ways of getting around degeneracies with other cosmological
parameters\cite{me1,me2,me4}.
The recent improvements in the CMB data sets available
(and particularly the availability of CMB polarisation data) will provide
crucial improvements. For example, we have recently shown\cite{me6} that
once data from the Planck Surveyor satellite is available, one will be
able to measure $\alpha$ to at least 0.1 percent accuracy.

\section{The Future}

Sespite tantalising hints the
currently available data is still inconclusive. What is needed to get a
definitive answer? Recent efforts have focused on the astrophysical side: a number of groups (including
our own) have started programmes aiming to obtain further measurements using
quasar data, or developing entirely new techniques, and there is reason to
hope that the situation will soon be much clearer. However,further key developments will be required. Among these, three spring to mind:

First, one would like to have a post-Planck, CMB polarisation-dedicated
satellite experiment. It follows from our recent CMB analysis and
forecasts\cite{me6} that the Planck Surveyor will measure CMB temperature
with an accuracy that is very close to the theoretically allowed precision
(technically, this is called the cosmic variance limit), but its polarisation
measurements will be quite far from it---this is because the ways one optimises
a detector to measure temperature or polarisation are quite different, and in
some sense mutually incompatible. On the other hand, we have also
shown\cite{me5,me6} that CMB polarisation alone contains much more cosmological
information than CMB temperature alone. Hence an experiment optimised for
polarisation measurements would be most welcome.

Second, one needs further, more stringent local tests of Einstein's Equivalence
Principle. This is the cornerstone of Einstein's gravity, but all theories with
additional spacetime dimensions violate it---the issue is not whether or not
they do, it's at what level they do it (and whether or not that level
is within current
experimental reach). In the coming years some technological developments are
expected which should help improve these measurements, not only in the lab
but also at the International Space Station and even using dedicated satellites
(such as $\mu$SCOPE, GG and STEP). Either direct violations will
be detected (which will be nothing short of revolutionary) or, if not, it will
drastically reduce the list of viable possibilities we know of, indicating that
the `theory of everything' is very different from our current
naive expectations.

Third, one needs tests of the behaviour of gravity. Because it is so much weaker
than the other forces of nature, it is quite hard to test the behaviour of
gravity, and surprisingly little is known about it on many interesting
scales. There has been a recent surge of interest in laboratory tests on
small scales (and further improvements are expected shortly), but experiments
that can test it on very large (cosmological) scales are still {\it terra
incognita}. This is relevant because in theories with
additional spacetime dimensions gravity is non-standard on large enough
and/or small enough scales. Furthermore, a non-standard large-scale behaviour
could also be an alternative to the presence of dark energy (a topic of
much recent interest, but beyond the scope of this article).

Apart from the experimental and observational work, there are deep
theoretical issues to be clarified. Noteworthy among these
is the question as to whether all
dimensionless parameters in the final physical `theory of everything'
will be fixed by consistency conditions or if some of
them will remain arbitrary. Today this is a question of belief---it does
not have a scientific answer. By arbitrary I mean in this context that a
given dimensionless parameter assumed its value in the process of the
cosmological evolution of the universe at an early stage of it. Hence, with
some probability, it could also have assumed other values,
and it could possibly also change in the course of this evolution.

\section{So What Is Your Point?}

Physics is a logical activity, and hence (unlike other intellectual pursuits),
frowns on radical departures. Physicists much prefer to proceed by
reinterpretation, whereby elegant new ideas provide a sounder basis for what
one already knew, while also leading to further, novel results with at
least a prospect of testability. However, it is often not easy to see how
old concepts fit into a new framework.
How would our views of the world be changed if decisive evidence is
eventually found for extra dimensions and varying fundamental constants?

Theories obeying the Einstein and Strong Equivalence Principles
are metric theories of gravity\cite{7}. In such theories the spacetime
is endowed with a symmetric metric, freely falling bodies follow
geodesics of this metric, and in local freely falling frames the
non-gravitational physics laws are written in the language of special
relativity. If the Einstein Equivalence Principle holds,
the effects of gravity are equivalent to the effects of living
in a curved spacetime. The Strong Equivalence Principle contains the
Einstein Equivalence Principle as the special case where local
gravitational forces (such as Cavendish experiments or
gravimeter measurements, for example)
are ignored. If the Strong Equivalence Principle is strictly valid,
there should be one and only one gravitational field in the universe,
namely the metric.
Varying non-gravitational constants are forbidden by General Relativity and
(in general) all metric theories. A varying fine-structure constant will
violate the Equivalence Principle, thus signalling the breakdown of
4D gravitation as a geometric phenomenon. It will also
reveal the existence of further (as yet undiscovered) gravitational fields
in the universe, and may be a
very strong proof for the existence of additional spacetime dimensions.
As such, it will be a revolution---even more drastic
than the one in which Newtonian gravity became part of Einsteinian gravity.
And while not telling us (by itself) too much about the `theory of
everything', it will provide some strong clues about what and where to
look for.

\section*{Acknowledgments}
This work was done in collaboration with P. Avelino, J.
Brinchmann, D. McIntosh, A. Melchiorri, A. Moorwood, J. Oliveira, G. Rocha, G. Rudnick, R. Trotta and P. Viana  and supported by 
grants FMRH/BPD/1600/2000, ESO/FNU/43753/2001 and CERN/FIS/43737/2001
from FCT (Portugal).
Numerical work was performed on COSMOS, the Altix3700
owned by the UK
Computational Cosmology Consortium, supported by SGI, HEFCE and PPARC.
I'm also grateful to J. Garcia-Bellido, M.
Murphy and R. Thompson for useful discussions and comments.

\end{document}